\def\comment#1{}

\newcommand{\beg}{\begin{eqnarray}}
\newcommand{\eee}{\end{eqnarray}}

\documentclass[prl,twocolumn,aps,epsf]{revtex4}
\usepackage{dcolumn}
\usepackage{amsmath}
\usepackage{graphicx}%
\def\cm#1{}

\begin{document}
\title{ Fractional-flux vortices and spin superfluidity in triplet superconductors
}
\author{
Egor Babaev}
\affiliation{LASSP, 
Cornell University,  Ithaca, NY 14853-2501 USA
\\ Department of Physics, Norwegian University of Science and Technology, N-7491 Trondheim, Norway 
}
\begin{abstract}
We discuss a novel type of  fractional
flux vortices along with  integer flux vortices
 in Kosterlitz-Thouless transitions
in a triplet superconductor. We show that under certain
conditions a spin-triplet  superconductor should exhibit 
a novel state of  {\it spin superfluidity} without  superconductivity.
\end{abstract}
\maketitle
\newcommand{\la}{\label}
\newcommand{\aaa}{\frac{2 e}{\hbar c}}
\newcommand{\Pfaff}{{\rm\, Pfaff}}
\newcommand{\kA}{{\tilde A}}
\newcommand{\G}{{\cal G}}
\newcommand{\cP}{{\cal P}}
\newcommand{\M}{{\cal M}}
\newcommand{\E}{{\cal E}}
\newcommand{\btd}{{\bigtriangledown}}
\newcommand{\W}{{\cal W}}
\newcommand{\X}{{\cal X}}
\renewcommand{\O}{{\cal O}}
\renewcommand{\d}{{\rm\, d}}
\newcommand{\bfi}{{\bf i}}
\newcommand{\e}{{\rm\, e}}
\newcommand{\bfx}{{\bf \vec x}}
\newcommand{\bfn}{{ \vec{\bf  n}}}
\newcommand{\bfs}{{\vec{\bf s}}}
\newcommand{\bfE}{{\bf \vec E}}
\newcommand{\bfB}{{\bf \vec B}}
\newcommand{\bfv}{{\bf \vec v}}
\newcommand{\bfU}{{\bf \vec U}}
\newcommand{\bfp}{{\bf \vec p}}
\newcommand{\f}{\frac}
\newcommand{\bfA}{{\bf \vec A}}
\newcommand{\non}{\nonumber}
\newcommand{\be}{\begin{equation}}
\newcommand{\ee}{\end{equation}}
\newcommand{\ba}{\begin{eqnarray}}
\newcommand{\ea}{\end{eqnarray}}
\newcommand{\bastar}{\begin{eqnarray*}}
\newcommand{\eastar}{\end{eqnarray*}}
\newcommand{\half}{{1 \over 2}}
Superconductors with triplet pairing allow for a rich variety
of topological defects and phase transitions  \cite{co5,JETP,book,samokh,mineev,voloviksplit}.
For example in $p$-wave superconductors there exist
different realizations of 
fractional vortices such as: the half-quantum vortex (the Alice string),
which is a vortex where phase changes by $\pi$ and spin is reversed when we go
around the vortex core, fractional vortices trapped on a grain
boundary, fractional flux trapped by twisted wire and other 
possibilities (for an excellent review see \cite{co5,book}).
Another subject which was intensively
studied is  the split phase transitions 
in triplet systems (e.g. in the presence of disorder or magnetic field)
 \cite{voloviksplit,samokh,book}. There have also been  
particularly interesting studies of split transitions 
in a neutral $p$-wave superfluid in connection with
thin films of liquid ${}^3He$ \cite{korshunov,stein}.
Also the related
 questions of  various partial symmetry breakdowns
are relevant and were studied in spin-1  Bose condensates in optical
traps \cite{deml1} and bilayer Quantum Hall systems \cite{deml2}.
In this Letter we discuss  the effect of flux fractionalization
due to spin-orbit coupling, as well as  the appearance of neutral vortices
 in triplet superconductors and 
discuss its influence on the Kosterlitz-Thouless (KT) transitions.
Based on  topological arguments we predict the existence of a novel type
of ordering in spin-triplet system - the spin-superfluid nonsuperconducting state.

We consider a model spin triplet superconductor
similar to one considered in \cite{samokh}
with order parameter 
$\Psi_{a}({\bf r}) = \sqrt{n} ({\bf x}) \zeta_a ({\bf x})$
where  $(a=1,0,-1)$ and $\zeta$ is
a normalized spinor $\zeta^\dagger \cdot \zeta=1$.
If one neglects  mixed gradient terms,
the free  energy density 
 can be written as \cite{samokh,ho,mineev,rosenstein}:
\beg
F&=& 
\f{\hbar^2}{2 M} (\nabla \sqrt{n} )^2 +
 \f{\hbar^2 n }{2 M } \left|\left(\nabla  +  i \f{2e}{\hbar c}{\bf A}\right)\zeta_a\right|^2
 - \mu n \nonumber \\&+&\f{n^2}{2} \left[ c_0 +c_2 <{\bf F}>^2\right] +\f{{\bf H}^2}{8\pi}, 
\label{neu}
\eee
where  $<{\bf F}> =\zeta_a^*{\bf F}_{ab}\zeta_b$ is the spin.
The presence of a crystal
lattice makes the mass $M$ a symmetric tensor $M_{ij}$.
Spinors  are related
to one another
 by gauge transformation $e^{i\theta}$ and spin rotations
${\cal U}(\alpha, \beta, \tau)$$=$$e^{-iF_{z}\alpha} e^{-iF_{y}\beta}
e^{-iF_{z}\tau}$,  where $(\alpha, \beta, \tau)$ are the
Euler angles. The matrices $\{F_x,F_y,F_z\}$ are:
\begin{eqnarray}
\left\{\left( \begin{array}{ccc} 0& \f{1}{\sqrt{2}}&0 \\ \f{1}{\sqrt{2}}&0& \f{1}{\sqrt{2}}
\\0& \f{1}{\sqrt{2}}&0\end{array} \right),   
\left( \begin{array}{ccc} 0& \f{-i}{\sqrt{2}}&0 \\ \f{i}{\sqrt{2}}&0& \f{-i}{\sqrt{2}}
\\0& \f{i}{\sqrt{2}}&0\end{array} \right),
\left( \begin{array}{ccc}1&0&0 \\0&0&0\\0&0&-1\end{array} \right)\right\}
\end{eqnarray}
The ground state structure of $\Psi_{a}({\bf r})$ can be found 
by minimizing the energy with fixed particle number \cite{samokh,ho}. 
When  $c_{2}<0$ the energy is minimized by 
$<{\bf F}>^2=1$ and the ground state spinor and density are \cite{ho}
\begin{eqnarray}
\zeta& =& 
 e^{i(\theta-\tau)} \left( \begin{array}{c}
e^{-i\alpha}{\rm cos}^{2}\frac{\beta}{2}\\
\sqrt{2} {\rm cos}\frac{\beta}{2}{\rm sin}\frac{\beta}{2}
 \\ e^{i\alpha}{\rm sin}^{2}\frac{\beta}{2} \end{array} \right); 
n = \frac{\mu}{c_0+c_{2}}
\label{ferrodensity} \end{eqnarray}
In this case there 
exists an  equivalence  between the gauge transformation  
associated with $\theta$ and 
spin rotations associated with $\tau$ and  the vacuum manifold is  $SO(3)$ \cite{ho,book,samokh}.
The equation for the supercurrent in this case depends
 on phase gradients and also on spin texture (the Mermin-Ho relation):
\beg
{\bf J} &=& \f{i \hbar e n}{M}\left( \zeta_a^* \nabla \zeta_a -\zeta_a \nabla \zeta_a^*  \right)
-\f{4 e^2 n}{M c }{\bf A}= \nonumber \\
&&  \f{ 2\hbar e n}{M}[\nabla (\theta-\tau) - \cos\beta \nabla \alpha] 
-\f{4 e^2 n}{M c }{\bf A}
\la{cur}
\eee
In \cite{p} it was discussed that (\ref{neu})
can be rewritten in the form of an extended
Faddeev model \cite{nature} related to that derived for 
two-gap superconductors \cite{we,rema}.
That is, 
introducing notations  $\nabla_i = \f{d}{dx_i}$, 
$\bfs =  (\sin\beta \cos\alpha, \sin\beta\sin\alpha, \cos\beta ) $, 
$ \vec{\cal C} = \f{M}{e n}{\bf J}$
and separating variables, the free energy density
(\ref{neu}) can be rewritten in terms of 
only gauge-invariant variables as
\beg
 &&F=
\f{\hbar^2 (\nabla \sqrt{n} )^2}{2 M}+
\f{\hbar^2 n (\nabla \bfs)^2 }{4 M} 
 - \mu n +\f{n^2\left[ c_0 +c_2 \right]}{2}+
 \nonumber \\ &&
 \f{n\vec{\cal C}^2}{8M}
+\frac{\hbar^2 c^2}{128 \pi e^2}\left(\f{[\nabla_i {\cal C}_j -\nabla_j 
{\cal C}_i]}{\hbar} -\bfs \cdot \nabla_i
\bfs \times \nabla_j\bfs
\right)^2 
\label{fa}
\eee

Let us now allow for a spin-orbit coupling breaking the $O(3)$ symmetry.
In the simplest case  one should add to (\ref{fa})
a term \cite{samokh}:
\beg
F_{so}=\gamma_{ii}\bfs^2+ \gamma_{ij}s_is_j
\label{so}
\eee
The  case when there is 
only one nonzero coefficient
 $\gamma_{zz}<0$ corresponds to the
``easy-axis" situation, then $O(3)$ spin symmetry 
is broken down to a point.
When  $\gamma_{zz}>0$ we have the ``easy plane" case
and the symmetry is broken down to a circle.
Adding the  terms
$\gamma_{xx}=\gamma_{yy}$
still leads to an easy-plane or an easy-axis
case.
However, in the general case
there should be higher order
corrections e.g. fourth  order  in $s_i$.
Such corrections shift the ground state from easy-axis or
easy plane and
the energetically preferred 
state of spin direction can be  a circle on 
the unit sphere as shown on Fig. 1.
Such a situation is known  in many
magnetic systems \cite{magnetic}. 
Let us now examine the consequences of the 
spin-orbit coupling for the vortices of $S^1\rightarrow S^1$ map.
Let us consider the case when spin-orbit coupling breaks 
spin rotation symmetry
down to an arbitrary  circle on $S^2$ characterized by $\beta=\beta_0$ (see Fig. 1)
(we consider the range $0\leq \beta_0\leq\pi$).
\begin{figure}[h]
\center{\includegraphics*[scale=0.18]{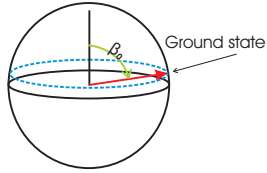}}
\caption{Ground state of $\bfs$ for $\beta_0 \ne \pi/2$  }
\end{figure}
One can observe  that when $\beta=\beta_0=const$ we have
$\bfs \cdot \nabla_i
\bfs \times \nabla_j\bfs \propto \sin\beta[\nabla_i\beta\nabla_j\alpha-\nabla_j\beta\nabla_i\alpha]= 0$.
In this case, in the London limit, the eq. (\ref{fa}) becomes:
\beg
&&F=\f{\hbar^2 n}{4 M}\sin^2\beta_0 (\nabla \alpha)^2 + \nonumber \\
&& \f{\hbar^2 n}{2M} \Biggl(\nabla (\theta-\tau) - \cos\beta_0 \nabla \alpha 
-\f{2e}{\hbar c}{\bf A}\Biggr)^2 +\f{{\bf H}^2}{8\pi}
 \label{fa2}
\eee 
From (\ref{fa2}) it follows that the system allows the following vortices of $S^1 \rightarrow S^1$ map:
(i) A vortex where $\theta$ (or $\tau$) changes by $2\pi$ around the core.
This is the analogue of Abrikosov vortex in ordinary superconductors and 
it carries one flux quantum. 
(ii) A vortex where $\alpha$  changes by
 $2\pi$ around the core. This vortex has
 a neutral vorticity  with the stiffness $\f{\hbar^2 n}{2M}\sin^2\beta_0$
described by the first term in (\ref{fa2}) and also
 there is a supercurrent around this vortex due to
 Mermin-Ho relation, which is  given by
\beg
{\bf J} = -\f{ 2\hbar e n}{M} \cos\beta_0\nabla \alpha 
-\f{4 e^2 n}{M c }{\bf A}
\la{cur2}
\eee
From here it follows that the magnetic flux 
of this vortex is determined solely by spin-orbit coupling 
and is an arbitrary fraction of magnetic flux quantum:
\be
\Phi=\oint_{\sigma} {\bf A} d{\bf l}=- \cos \beta_0 \Phi_0,
\la{flux}
\ee
where $\sigma$ is a path around the core at a distance much larger
than $\lambda$ and $\Phi_0$ is the magnetic flux quantum.
A special situation appears in  the ``easy-plane" case,
when $\beta_0=\pi/2$, 
and a vortex in spin degree of freedom does not carry magnetic flux.

Let us now discuss the consequences of the above
topological excitations to phase diagram of a quasi-two-dimensional
model spin-triplet superconductor where the $O(3)$ symmetry
is  broken to $O(2)$. We consider the case when  this
symmetry is broken strongly and
restrict discussion to the range $0\leq \beta \leq \pi/2$,
with the main focus on the case $\beta=\pi/2$.
Thus for simplicity we do not consider in this Letter
the case of two degenerate minima with $\beta=\beta_0$ and $\beta=\pi-\beta_0$).
From now on we discuss a quasi-two dimensional system of thickness $d$
where the effective magnetic field penetration length $\lambda=\lambda_P$
 is related to Ginzburg-Landau penetration length $\lambda_{GL}$ via
the Pearl expression \cite{1}
$\lambda_P=\lambda_{GL}^2/d$. 

We begin with the simplest limit when the magnetic field penetration length $\lambda$
is larger than the sample size and $\beta_0=\pi/2$ (the easy plane case).
In such a situation we can neglect 
coupling to the vector potential in (\ref{fa2}).
 Then we can observe that 
vortices where   ($\Delta\theta = \oint dl \nabla \theta = 2\pi$,$\Delta\alpha = \oint dl \nabla \alpha =0$)
are mapped onto vortices in  the ordinary one component  $U(1)$
model with stiffness $J_\theta= \f{\hbar^2 n}{M}$,
while the vortices 
($\Delta\alpha = \oint dl \nabla \alpha = 2\pi$,$\Delta\theta = \oint dl \nabla \theta = 0$)
are mapped onto vortices in a  $U(1)$-model with stiffness
$J_\alpha= \f{\hbar^2 n}{2M}$. 
Thus  in the limit $\lambda \to \infty$
there are no corrections to vortex interaction due
to the Meissner effect, and
the system    
undergoes two KT phase transitions at the temperatures: 
\beg &&
T_{KT(\theta)}^{\lambda \to \infty,\beta_0=\pi/2}
=\f{\pi}{2}
\f{\hbar^2 n(T_{KT (\theta)}^{\lambda \to \infty,\beta_0= \pi/2})}{M} \nonumber \\
&&T_{KT (\alpha)}^{\lambda \to \infty,\beta_0= \pi/2}
=
\f{\pi}{4}
\f{\hbar^2 n(T_{KT(\alpha)}^{\lambda \to \infty,\beta_0= \pi/2})}{M} 
\label{elem}
\eee
(equations for $T_{KT}$ should be solved
self-consistenly with the equations for gap modulus, see e.g. the review \cite{ijmp}).
This splitting of the phase transitions
corresponds to separate onset of quasi-long-range order
in phases $\theta$ and $\alpha$.
\comment{
However the case when $0<\beta_0 < \pi/3$
is very different. That is, in a KT transition the system
can excite vortices 
  ($\Delta\theta  = 2\pi$,$\Delta\alpha =2 \pi $)
and antivortices 
  ($\Delta\theta  = -2\pi$,$\Delta\alpha =-2 \pi $)
which are composite, yet have lower energy
comparing to the elementary vortices 
($\Delta\theta  = \pm 2\pi$,$\Delta\alpha =0 $)
and ($\Delta\theta  =0 $,$\Delta\alpha = \pm 2 \pi $).
This is because the composite vortices have smaller second term in (\ref{fa2}).
Thus when $0<\beta_0 < \pi/3$ the system undergoes 
a phase transition of unpairing of 
composite vortices and antivortices 
 ($\Delta\theta  =\pm 2\pi$,$\Delta\alpha = \pm 2 \pi $)
at 
\beg
&&T_{KT(\alpha+\theta)}^{\lambda \to \infty,0<\beta_0 < \pi/3}= \nonumber \\
&&
\pi \f{\hbar^2 n(T_{KT(\alpha+\theta)}^{\lambda \to \infty,0<\beta_0 < \pi/3})}{M} \Bigl[1-\cos\beta_0-\f{\sin^2\beta_0}{4}
\Bigr]
\eee
Since it is the composite vortices which participate
in the transition $T_{KT(\alpha+\theta)}^{\lambda \to \infty,0<\beta_0 < \pi/3}$,
we have a constrain that  during this transition, 
individually, the phase $\alpha$ and $\theta$
disorder but the difference $(\alpha-\theta)$ of these phases remains quasi-ordered.
The system becomes completely disordered
only at a higher temperature
which corresponds to liberation of 
elementary vortices and antivortices (\ref{elem}).}

The  regime 
of short penetration  length 
is however principally different and more interesting.
In this limit the interaction 
which is mediated by charged current 
between vortices
and antivortices  ($\Delta\theta = \pm 2\pi$) is exponentially screened
at the length scale $\lambda$ and such vortices do not undergo a
KT transition. Thus the variable $\theta$ is disordered and the system
is not superconducting. However  in this limit  the 
vortices with  ($\Delta\alpha = \pm 2\pi$) still have a long range 
interaction mediated by the neutral mode and characterized by stiffness:
\be 
J^{\alpha}_c=\sin^2\beta_0  \f{\hbar^2 n}{2M}
\ee
(as follows from the first term in (\ref{fa2})). Thus 
the ``spin vortices" in the regime
$\pi/3<\beta_0\leq  \pi/2$  can undergo a true KT transition at 
the temperature 
\be
T_{KT}^{SSF}=\f{\pi}{4} \sin^2\beta_0\f{\hbar^2 n(T_{KT}^{SSF })}{M}
\la{pt}
\ee
this phase transition is associated with establishment of
quasi-long range order in the phase variable $\alpha$.
Note that in this state the superconducting phase $\theta$
is still disordered so $T_{KT}^{SSF}$ is
{\it not} a superconducting phase transition
but is a  transition to  a ``{\it spin superfluid state}".
Let us emphasise that the vortices  with ($\Delta\alpha = \pm 2\pi$)
have nontrivial contribution in the second term in (\ref{fa2})
and also carry a fraction of flux quantum (\ref{flux}).

We should observe that the situation 
when $0<\beta_0<\pi/3$
is very special. That is,  in the short-$\lambda$ limit 
the long range interaction of
vortices is mediated by the neutral current 
(coming from the neutral mode 
described by the first term in (\ref{fa2}))
and besides that there is a finite energy contribution  from
 the charged current (described by the 
the second term in (\ref{fa2})). Although the later
contribution is irrelevant for vortex interaction
(in Coulomb gas mapping it plays
the same role as the core energy), however
the charged mode is important in the following respect:
In the regime $0<\beta_0<\pi/3$ 
the {\it composite} vortices ($\Delta\alpha =  \pm 2\pi,\Delta\theta = \pm2\pi$)
have smaller energy than {\it elementary} vortices
($\Delta\alpha = \pm 2\pi,\Delta\theta =0$)
since the former vortices carry a smaller fraction of
magnetic flux quantum and thus have smaller
kinetic energy of charged supercurrent.
Thus when  $0<\beta_0<\pi/3$
there will be a KT transition of the composite 
vortices  ($\Delta\alpha =  2\pi, \Delta\theta = 2\pi$)
and antivortices  ($\Delta\alpha =  - 2\pi, \Delta\theta = -2\pi$)
at the temperature (\ref{pt}). This is a rather unique feature,
rooted in the nontrivial coupling of phases by vector potential in 
the triplet superconductor,
when a KT transition is governed by topological defects with 
high topological charge.
Note that in this regime 
the system allows also ``purely charged" 
integer flux vortices (that is, without neutral superflow)
 ($\Delta\alpha =0, \Delta\theta =\pm 2\pi$).
These vortices have only screened interaction. In
this regime they are always liberated and
disorder superconducting phase. Thus,
also in the regime $0<\beta_0 < \pi/3$ the 
phase transition (\ref{pt}) is  a transition
to a nonsuperconducting spin-superfluid state 
in spite it is mediated by composite vortices
which have topological windings in both phases
($\Delta\alpha =  \pm 2\pi, \Delta\theta = \pm 2\pi$).

\begin{figure}[h]
\center{\includegraphics*[scale=0.4]{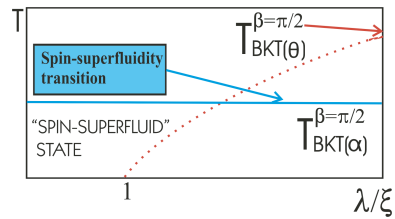}}
\caption{\label{pd} A schematic phase diagram of the system
in the simplest regime $\beta_0=\pi/2$. The temperature of onset of
superconductivity, at finite penetration length 
is a characteristic temperature of a washed out crossover.
}
\end{figure}

Let us now discuss  qualitatively  regimes of intermediate
penetration length.
In a general case of large but finite $\lambda$,
the interaction of vortices and antivortices, 
which is mediated by charged current, receives
correction due to Meissner screening (for a detailed review
and citations see \cite{1}). Then,
if $\lambda$ is sufficiently large, a gauged purely $U(1)$ system,
althought, strictly speaking,
does not display  a KT transition into a superconducting state,
 is however believed to
undergo a ``would be" KT crossover \cite{1}. So,
$T_{KT}$ at finite $\lambda$ is transformed
into a characteristic temperature of a broader
crossover and also is shifted down  comparing to the 
situation when there is no Meissner screening \cite{1}.
In the system (\ref{fa2}), at finite-$\lambda$,  the vortices  ($\Delta\theta = \pm 2\pi$)
interact via {\it charged} mode with the effective stiffness:
\be
J^\lambda_\theta(T)=
\f{\hbar^2 n(T)}{M}
\label{jc2}
\ee
while the vortices with ($\Delta\alpha = \pm 2\pi$)
interact by means of both the  {\it neutral} spin mode with the effective stiffness 
\be
J^\lambda_{\alpha n}(T)= \sin^2\beta_0\f{\hbar^2 n(T)}{2M} 
\ee
and with charged mode, which
 effective stiffness  in the case $\pi/3<\beta_0 \leq \pi/2$
is 
\be
J_{\alpha c}^{\lambda,\pi/3<\beta_0 \leq \pi/2}(T)=\cos^2\beta_0 \f{\hbar^2 n(T)}{M}
\label{jc2}
\ee
%
When $0<\beta_0<\pi/3$ the relevant topological 
excitation is the composite vortex ($\Delta \alpha=2\pi,\Delta\theta=2\pi$)
for which
\be
J_{\alpha c}^{\lambda,0<\beta_0<\pi/3}(T)=[1-\cos\beta_0]^2\f{\hbar^2 n(T)}{M}
\label{jc22}
\ee
Let us now consider the easy-plane case $\beta_0=\pi/2$.
In type-I limit, as discussed above, there is only spin-superfluid
KT phase transition while superconductivity sets in only
at $T \to 0$. When we increase penetration length the
system undergoes a washed out superconducting KT crossover  \cite{1}
 at some nonzero temperature. 
In the extreme limit $\lambda \to \infty$
we arrive at the above limit of 
two true KT transitions
and an  intermediate state where
there is no quasi-long range order in spin degrees
of freedom while the system is superconducting.
The case of arbitrary $\beta_0$ and $\lambda$
appears to be a very interesting subject for 
numerical simulations.
A schematic phase diagram in the simplest case $\beta_0 = \pi/2$
is given on Fig. \ref{pd}.

Let us discuss the
 physical interpretation of the ``spin-superfluid" state.
In this 
state  a quasi-long-range order is retained in the spin angle $\alpha$.
We observe that all three complex components in Eq. (\ref{ferrodensity}) have 
a prefactor $e^{i\theta}$. Thus in the spin-superfluid state,
 all three phases of the spin components  are disordered.
Consequently, there is no dissipationless supercurrent (\ref{cur}).
However the system retains hidden order, that is,
taking the difference of the total phases of the first and 
third components, the contributions from the phase $\theta$
is cancelled, while the contribution of the ordered phase $\alpha$ is
not.  That means that there is a quasi-long-range order in  a
composite order parameter. Such a composite order parameter
can be identified though the principle that a squared modulus of its gradient 
should give the kinetic term of the composite neutral mode
in the GL functional after separation of variables 
(\ref{fa2}). This term is the following
combination of the fields $\zeta_a$
which is decoupled from the vector potential 
but is gauge invariant:
$(\hbar^2n /2M) [\nabla \zeta_a \nabla\zeta_a^*+(1/4)( \zeta_a^* \nabla \zeta_a -\zeta_a \nabla \zeta_a^* )^2]|_{\beta=\beta_0}$
This corresponds to the following order parameter
for the composite neutral mode:
\be
\Xi=|\Xi|e^{i\alpha}=\sqrt{n|\zeta_1\zeta_{-1}|}e^{i\alpha}=\sqrt{\frac{n}{2}}\sin{\beta_0}e^{i\alpha}.
\ee
In the hydrodynamic limit the spin superfluid velocity is given
by ${\bf V}_S=  \f{ \hbar n}{2M} \sin^2\beta_0 \nabla \alpha$. 
This expression  has a physical 
interpretation: in the easy-axis limits (when $\beta_0=0$ or $\beta_0=\pi$)
there is no neutral $U(1)$ symmetry  and we find that also ${\bf J}_S=0$.
On the other hand in the easy-plane case when $\beta_0=\frac{\pi}{2}$ 
we have $|\Psi_{1}|=|\Psi_{-1}|=|\Psi_{0}|/\sqrt{2}$. We note
that $\Psi_0$
is neutral with respect to $\alpha$ and thus does not participate
in spin superfluidity. This gives a natural
explanation for the extra factor $1/2$  in front
of the phase gradient in the expression for ${\bf V}_S$ 
in the easy-plane situation, compared  to 
the similar coeffcient in front of $\nabla \theta$
 in the expression for $(1/2e ){\bf J}$ given by eq. (\ref{cur}). 

\begin{figure}[h]
\center{\includegraphics*[scale=0.2]{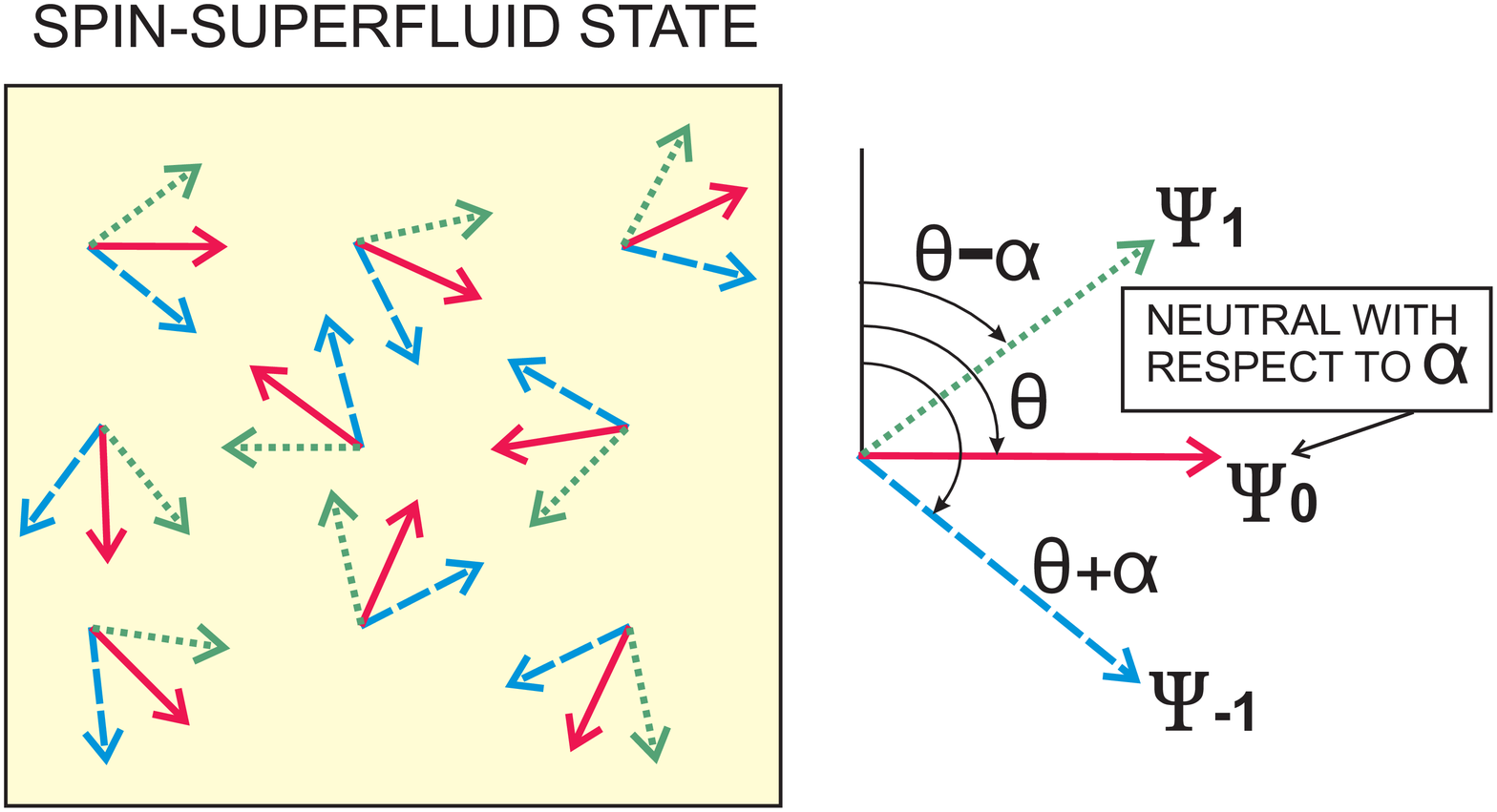}}
\caption{\label{pha} A schematic illustration of 
the configuration of phases of individual spin components
in different points in space
in the spin-superfluid state.}
\end{figure}

%
The configuration  of the phases of the three complex components
of the order parameter
in the spin-superfluid state is 
illustrated in Fig. \ref{pha}.

Possible experimental probes of the spin-superfluid state are:
({\it i}) When there are corrections shifting spin from the easy plane,
the vortices in the $\alpha$-phase  carry
a fraction of flux quantum. The intrinsic flux noise of a 
      dc SQUID is about $10^{-6}$ $\Phi_0/\sqrt{Hz}$ 
     which  makes flux-noise (FN) measurements sensitive
     to spin-superfluid transition even if  a tiniest ``shifting correction"
     is present.
An observation of a KT transition in the spin-superfluid state  in an
    FN experiment  should allow to distinguish it 
     from a superconducting transition:
     Because of the existence of the neutral mode,
the transition should be sharp and a true jump in superfluid
            density should be recoverable from flux-noise data, 
            in contrast to similar measurements of wider KT crossovers in ordinary
            superconductors. 
({\it ii}) 
Upon the transition to spin superfluid state
            a ``true" superconductivity is absent. 
            So after detection of the
transition by a FN measurement, in subsequent measurements of the
conductivity or application of an external field, the system will
not behave as a superconductor
  ({\it iii})   If  the corrections shifting the spin from the equator on $S^2$
             are absent - there nonetheless  is a possibility to observe the transition
             to the spin-superfluid state if 
            an  external field is applied. Then the
             Zeeman effect would result in  a term linear in $s_z$ in (\ref{fa}) thus
             shifting spins from equator on $S^2$, consequently the spin vortices
             will acquire a fraction of flux quantum 
and the transition will be detectable via FN measurement.

In conclusion, 
a triplet 
superconductor
is  a multicomponent 
charged system, where 
the vector potential is coupled
to phase variables in a  nontrivial way which results in the
existence of  neutral modes. 
The nontrivial coupling 
to vector potential of  the
components of the order parameter 
becomes particularly important in (quasi)-two dimensions where 
the system undergoes topological phase transitions.
We considered a simple model
of a  spin-1 superconductor
and have shown that topological 
considerations lead to  a novel  physical state:
the ``spin-superfluidity" 
when only an opposite superflow of two of three spin
components is allowed.
The results hold true in the simplest case of the easy-plan situation, 
however we extended the discussion to a general case when 
the ground state is shifted from the equator on $S^2$ which, 
as we have shown, leads to flux fractionalization.
We also pointed out that there is a range of parameters
when the quasi-long range order sets in 
only in the superconducting phase 
and pointed out an unusual KT transition mediated by
defects of high topological charge.

We thank A. Sudb{\o}, N. W Ashcroft, E. Demler, O. Eriksson, M. Katsnelson, S. Ktitorov,
M. Zhitomirsky
and especially G. E. Volovik,
D. Agterberg and S. Korshunov for many generous discussions.    
This work has been supported by STINT and Swedish Research Council,
 Research Council of Norway, Grant No.
157798/432 and National Science Foundation,
Grant DMR-0302347.
 

\begin{thebibliography}{99}
\bibitem{co5}
For a review of other examples
of flux fractionalization in triplet superconductors 
see
G.~E.~Volovik,
Proc.\ Nat.\ Acad.\ Sci.\  {\bf 97}, 2431 (2000)
see also 
I. A. Lukyanchuk, M. E. Zhitomirsky
Sup. Rev.{\bf 1}, 207 (1995)
\bibitem{book}
G. E. Volovik {\it The Universe in a Helium Droplet} Clarendon Press (2003)
\bibitem{JETP}
G.E. Volovik, L.P. Gor'kov  Sov. Phys. JETP Lett. {\bf 39} 647  (1984) 
\bibitem{voloviksplit}
R. Joynt, V.P. Mineev, E.G. Volovik, M.E. Zhitomirsky Phys. Rev. {\bf B 42}, 2014 (1990)
and references  therein.
\bibitem{samokh}
L.I. Burlachkov, N.B. Kopnin Sov. Phys. JETP {\bf 65} 630 (1987).
\bibitem{mineev}
V. P. Mineev, K.V. Samokhin 
{\it Introduction to Unconventional Superconductivity}
Taylor \& Francis (1999). 

\bibitem{korshunov}
S. Korshunov Sov. Phys. JETP {\bf 62} (1985) 301.
\bibitem{stein}
 Stein and Cross, Phys. Rev. Lett. { \bf 42} 504 (1979).
\bibitem{deml1}
E. Demler, F. Zhou Phys. Rev. Lett. {\bf 88} 163001 (2002),
A. Imambekov, M. Lukin,  E. Demler 
 Phys. Rev. A {\bf 68} 63602 (2003),
A. Kuklov, N. Prokof'ev and B. Svistunov  Phys. Rev. Lett. {\bf 92}, 030403 (2004) 


\bibitem{deml2}
E. Demler, C. Nayak, S. Das Sarma 
Phys. Rev. Lett. {\bf 86} 1853 (2001)
\bibitem{ho}
Tin-Lun Ho Phys. Rev. Lett. {\bf 81} 742 (1998); see also
T. Ohmi, K. Machida J. Phys. Soc. Jpn. {\bf 67}, 1822 (1998).
\bibitem{rosenstein}
A. Knigavko and B. Rosenstein Phys. Rev. Lett. {\bf 82}, 1261 (1999),
A. Knigavko, B. Rosenstein, and Y. F. Chen
Phys. Rev. B{ \bf 60}, 550 (1999);
B. Rosenstein, {\it et. al.} 
Phys. Rev. B {\bf 67}, 224507 (2003) .
\bibitem{p}
E. Babaev Phys. Rev. Lett. {\bf 88} 177002 (2002)
\bibitem{nature} 
L.D. Faddeev, A.J. Niemi, { Nature}
{\bf 387} (1997) 58;  
L. Faddeev, {\it Quantisation of Solitons},
preprint IAS Print-75-QS70, 1975.
\bibitem{we} 
E. Babaev, L.D. Faddeev, A.J. Niemi 
Phys.\ Rev.\ {B \bf 65}, 100512 (2002),
a similar model might be realized also in neutron stars E. Babaev Phys. Rev. D 
{\bf 70}  043001 (2004)
\bibitem{rema}
The models are equivalent 
in a simply connected space. Vortices 
of $S^1\to S^1$ map (which make the physical space
multiply connected) are very different in these systems.
\bibitem{magnetic} see e.g.
M. Colarieti-Tosti {\it et. al.}
Phys. Rev. Lett. {\bf 91}, 157201 (2003)
B. R. Cooper {\it et. al. }, Phys. Rev. {\bf 127}, 57 (1962).



\bibitem{ijmp}
E.~Babaev,
Int.\ J.\ Mod.\ Phys.\ A {\bf 16}, 1175 (2001)


\bibitem{1}
P. Minnhagen, Rev. Mod. Phys. {\bf 59} 1001 (1987). 
\end{thebibliography}
\end{document}